\documentclass[12pt,preprint]{aastex} 

\shorttitle{Multiple sources toward S140~IRS~1 }

\shortauthors{Trinidad et al.}

\begin{document} 
 
\title{Multiple Sources toward the High-mass Young Star S140~IRS~1}

\author{Miguel A. Trinidad\altaffilmark{1} 
Jos\'e M. Torrelles\altaffilmark{2},
Luis F. Rodr\'{\i}guez\altaffilmark{3},
Salvador Curiel\altaffilmark{4}  }

\email{trinidad@astro.ugto.mx} 

\altaffiltext{1}{Departamento de Astronom\'{\i}a, Universidad de Guanajuato,
Apdo. Postal 144, Guanajuato, Gto. 36240, M\'exico} 
\altaffiltext{2}{Instituto de Ciencias del Espacio (CSIC) and Institut 
d'Estudis Espacials de Catalunya, Facultat de F\'{\i}sica, Planta 7a, 
Universitat de Barcelona, Av. Diagonal 647, 08028 Barcelona, Spain}
\altaffiltext{3}{Centro de Radioastronom\'{\i}a y Astrof\'{\i}sica, (UNAM),
Apdo. Postal 3-72 (Xangari) 58089 Morelia, Michoac\'an, M\'exico}
\altaffiltext{4}{Instituto de Astronom\'{\i}a, UNAM, Apdo. Postal 70-264, D.F.
04510, M\'exico} 

\begin{abstract} 
 
S140 IRS1 is a remarkable source where the radio source at the center
of the main bipolar molecular outflow in the region is elongated perpendicular
to the axis of the outflow, an orientation opposite to that expected
if the radio source is a thermal jet exciting the outflow. 
We present results of 1.3~cm continuum  and H$_2$O maser emission
observations  made with the VLA in its A configuration toward this region.
In addition, we also present results of continuum observations
at 7~mm and re-analyse observations at 2, 3.5 and 6~cm (previously 
published). IRS~1A is detected at all wavelengths, showing an elongated 
structure. Three water maser spots are detected along the major
axis of the radio source IRS 1A.
We have also detected a new continuum source at 3.5~cm
(IRS 1C) located $\simeq 0\farcs6$ northeast of IRS 1A. The presence of
these two YSOs (IRS~1A and 1C) could explain the existence of the two
bipolar molecular outflows observed in the region. In addition, we
have also detected three continuum clumps (IRS~1B, 1D and 1E) located
along the major axis of IRS~1A. We discuss two possible models to
explain the nature of IRS~1A: a thermal jet and an equatorial wind.

\end{abstract} 
 

\section{Introduction} 
 
The process of low-mass star formation is reasonably well understood.
The accepted model (e.g., Shu et al. 1987) requires a disk-YSO-outflow 
system (\textit{YSO: young stellar object}), and it has been supported
by theoretical and observational results (e.g. Evans, 1999 and references 
therein). In addition, bipolar molecular outflows associated with low-mass
stars are believed to be driven by their jets, that can be
observed at sub-arcsec scales and this constitute the best
evidence of collimation at the smallest scales now known
(e.g. Anglada 1996).
However, the process of high-mass star formation is not well understood
yet. In fact, although molecular outflows seem to be also common among 
high-mass stars (G\'omez et al. 1999, Zhang et al. 2001, Ridge \& Moore,
2001; Beuther et al. 2002, Shepherd 2005), there is still a deficit in
the detection of circumstellar disks and jets associated with massive
YSOs (e.g., Cepheus A HW2: Rodr\'{\i}guez et al. 1994; Patel et al. 2005; 
Curiel et al. 2006; IRAS~20126+4104: Cesaroni et al. 1999; 
Trinidad et al. 2005; Sridharan et al. 2005; AFGL 490: Schreyer et al.
2006; G24.78+0.08: Beltr\'an et al. 2006). In spite of these studies,
it is not clear whether molecular outflows in massive
YSOs are also driven by collimated jets as low-mass stars do. Therefore,
studies of individual high-mass YSOs are very important to address these 
issues, but their identification is more difficult than in the case of
low-mass objects. Firstly, the spatial density of massive objects is smaller
and we need to go to greater distances to increase the known sample and
build statistical significance, and second, they are formed in
dense clusters and sub-arcsecond angular resolution observations are required
to isolate them.

The star-forming region, S140~IRS, is located in the molecular cloud
L1204 at a distance of 910 pc (Crampton \& Fisher, 1974). In this
region, there is a group of at least three ZAMS B stars (IRS~1,
IRS~2, IRS~3; Beichman et al. 1979), with a total luminosity of
$\sim 3\times10^3$~L$_\sun$  (Lester et al. 1986), with IRS~1
being the brightest one. In addition, a CO bipolar outflow
extended along the northwest-southeast direction has been observed
in the region by Hayashi et al. (1987). Minchin et al. (1993) found that
the blue and redshifted lobes of the CO bipolar outflow have position
angles of $\sim 160\degr$ and $\sim 340\degr$, respectively. On the
other hand, K band (2.0-2.3~$\mu$m) and H$_2$ observations have
revealed two bipolar outflows in the region (Preibisch \& Smith 2002
and Weigelt et al. 2002), one of them with an orientation similar to
the CO outflow (160/340$\degr$) and the other one in the 20/200$\degr$
direction. Both bipolar outflows seem to be centered on IRS~1. The region
S140~IRS has also been studied at the infrared and optical bands (Eiroa
et al. 1993), NH$_3$ lines (Verdes-Montenegro et al. 1989),
and radio continuum emission (Schwartz 1989, Evans et al. 1989).
More recently, Hoare (2006) has observed IRS~1 at 6 cm during three
epochs with MERLIN, and shows, through a very detailed analysis that it is
highly elongated in the northeast-southwest direction, and also proposes
that the radio continuum traces an ionized equatorial wind driven by
radiation pressure from the central star and oriented in the 
northeast-southwest direction,
perpendicular to the CO bipolar outflow. In addition, water maser emission
has also been detected toward the S140~IRS region (e.g. Tofani et al. 1995,
Lekht et al. 1993, Lekht \& Sorochenko 2001, Trinidad et al. 2003).

In this paper, we report and discuss new high angular resolution observations
of 1.3~cm continuum and water maser emission toward the S140~IRS region
carried out with the Very Large Array (VLA) in the A configuration. 
In order to present a full study of IRS~1, we have also analyzed
7~mm continuum observations (VLA data archive; see also Gibb \& Hoare, 2007)
and re-analyzed VLA centimeter
continuum  observations (previously published by Schwartz 1989 and
Tofani et al. 1995). We describe the observations in \S2 and present
the results in \S3. In Section \S4, we discuss the nature of the
multiple continuum sources that we detect in the region, while
our main conclusions are summarized in \S5.

\section{Observations} 

The observations toward the S140~IRS region were made with the VLA of the
National Radio Astronomy Observatory (NRAO)\footnote{
The NRAO is operated by Associated Universities Inc., under cooperative
agreement with the National Science Foundation.} in its A configuration
on 1999 June 29. We observed simultaneously 1.3~cm continuum and
water maser emission.
We used two different bandwidths, one of 25~MHz with seven 
channels for the continuum and another one of 3.125~MHz with 63
channels for the maser line emission. Both the right and left circular 
polarizations were sampled in the above two different bandwidths, which were
averaged in order to improve the sensitivity. The broad bandwidth for
the continuum observations was centered at 22285.080~MHz, while the
narrow bandwidth for the line observations was centered at the frequency
of the H$_2$O $6_{16} \rightarrow 5_{23}$ maser line (rest frequency
22235.080~MHz) with V$_{LSR}=-9.0$~km~s$^{-1}$, covering  from
$-28$ to 10~km~s$^{-1}$ in velocity. The absolute amplitude calibrator was
3C~286 with an adopted flux density of 2.51~Jy, while the phase calibrator
was B2021+614 with a bootstrapped flux density of $2.14\pm0.05$~Jy.
The water maser line and continuum data were reduced and calibrated using
the standard techniques using the NRAO AIPS software package. After the
first calibration, we searched the narrow bandwidth data for the spectral
channel with the strongest water maser emission and then its signal
was self-calibrated in phase and amplitude. We then cross-calibrated
the data applying the phase and amplitude corrections to both the narrow
and the broad bandwidths. In this way, the atmospheric and instrumental
errors at high frequencies on long baselines were removed (see for details
Reid \& Menten 1990, Torrelles et al. 1996) and the signal-to-noise 
ratio was improved.

We also used the data set at 7~mm from the VLA data archive
(see also Gibb \& Hoare, 2007). In addition,
we re-analyzed continuum observations at 2, 3.5 and 6~cm, which have
been previously published (see Table \ref{observations}).
All observations were carried out in the A configuration.
We recalibrated these data sets and made new contour maps using the current
procedures of AIPS. We used the physical parameters derived from these
new contour maps for the discussion in this paper. Contour maps at 
2 and 6~cm and their measured physical parameters are consistent with
those published before by Schwartz (1989). However, the flux density
estimated at 3.5~cm is different (about $50\%$ lower) to that reported by
Tofani et al. (1995), probably due to a typo.

\section{Observational Results}  
\label{result}

\begin{figure} 
\plotone{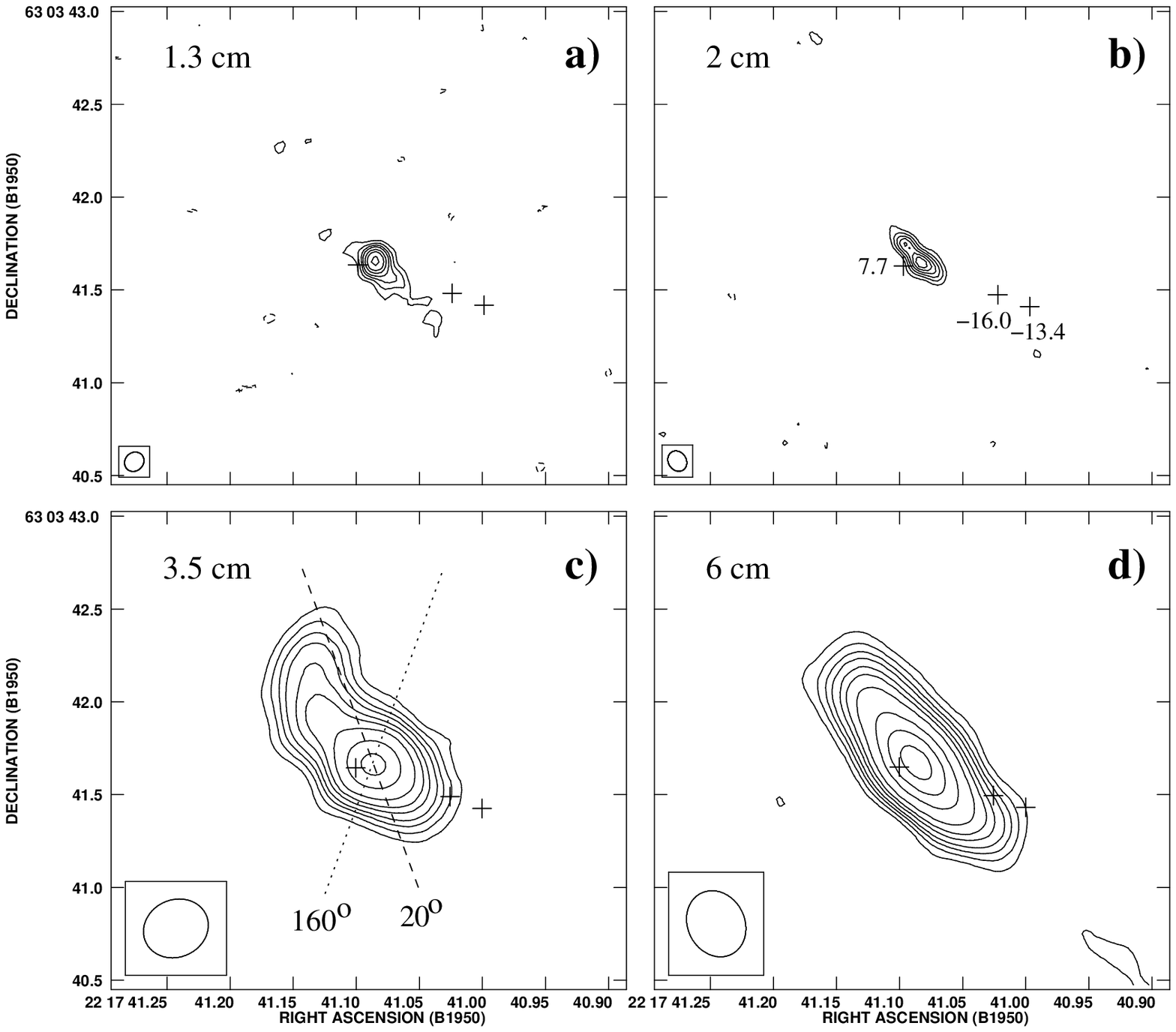}
\caption{Radio continuum contour maps showing consistently an elongated
structure for S140~IRS~1A at several wavelengths. Contours are 
--3,3,5,7,9,12,15 and 20 ({\it panel a:} 1.3~cm; beam
$0\farcs11\times0\farcs09$), 
--3,3,5,7,9,11, and 13 ({\it panel b:} 2~cm; beam 
$0\farcs12\times0\farcs10$), 
--3,3,5,7,9,12,15,20,30 and 40 ({\it panel c:} 3.5~cm; beam 
$0\farcs36\times0\farcs31$), and
--3,3,5,7,9,12,15,20,30,40 and 50 ({\it panel d:} 6~cm; beam 
$0\farcs37\times0\farcs31$) times 200, 170, 43 and 40~$\mu$Jy~beam$^{-1}$,
respectively, the rms noise of the maps.
In all maps the beam is shown in
the lower left corner. In order to make all continuum peaks of IRS~1A
coincide, we have applied an offset of ($\alpha$, $\delta$) =
($-0\rlap.^{s}0017$,  $-0\farcs007$),
($+0\rlap.^{s}0017$,  $+0\farcs009$), and 
($+0\rlap.^{s}0016$,  $+0\farcs014$)
to the position of IRS~1A at 2, 3.5 and 6~cm, respectively.
The crosses show the position of the water
masers detected in the region with their LSR velocity values (km s$^{-1}$)
indicated in panel b. The dashed and dotted lines in panel c indicate,
respectively, the axes of the H$_2$ jet and CO outflow seen on large scales.
} 
\label{IRS1}
\end{figure}

Figure \ref{IRS1} shows the contour maps of S140~IRS~1 at centimeter
wavelengths (1.3, 2.0, 3.5 and 6.0~cm). Given that other condensations
are also detected around S140~IRS~1, hereafter, we will refer to 
S140~IRS~1 as S140~IRS~1A or only IRS~1A for clarity. This source
appears spatially resolved at all wavelengths. The physical parameters
of the radio source 
IRS~1A (position, flux density, deconvolved angular size and position
angle) were obtained from elliptical Gaussian fits using the AIPS
task IMFIT and are given in Table \ref{flux}.
Contour
maps at 1.3 and 2~cm have similar angular resolution ($\sim 0\farcs1$)
and were made with natural- and uniform-weighting, respectively.
On the other hand, contour maps at 3.5 and 6~cm have an angular resolution
of $\sim 0\farcs3$ (Figure \ref{IRS1}). For the map at 3.6~cm a Gaussian taper
of 850~k$\lambda$ was used, while at 6~cm a ROBUST = --1
parameter (Briggs 1995) of the AIPS task IMAGR was used.

In all contour maps (Figure \ref{IRS1}), IRS~1A shows a general elongated
morphology in the northeast-southwest direction, similar to the
orientation of the bipolar outflow observed in H$_2$ toward IRS~1A
in the $20\degr/200\degr$ direction (Preibisch \& Smith 2002).
In addition, our contour map at 1.3~cm with angular resolution 
of $\simeq 0\farcs08$ (ROBUST = 0
parameter was used in order to optimize the compromise
between the signal-to-noise ratio and the angular resolution)
shows a second peak, at a level of 4$\sigma$, located about
$0\farcs13$ to the southwest from IRS~1A (Figure \ref{multi}). We will 
refer to this peak as IRS~1B. Furthermore, three other continuum peaks
(that we will refer as IRS~1C, IRS~1D and IRS~1E), are also observed
at 3.5~cm with an angular resolution of $0\farcs25$ (Figure \ref{multi}).
IRS~1C is located $\sim 0\farcs6$ to the northeast of IRS~1A (Figure
\ref{multi}) and is also detected at 2~cm with an angular resolution of
$\sim 0\farcs15$. IRS~1D and IRS~1E are located $\sim 0\farcs35$ to the
northeast and southwest of IRS~1A, respectively. Their main observed
parameters are given in Table \ref{flux}.

On the other hand, IRS~1A is detected at 7 mm (Figure \ref{Qband};
flux density and deconvolved size are given in Table \ref{flux})
while this is not the case for IRS~1B, 1C, 1D or 1E. Although the 7~mm
continuum emission is also elongated in the northeast-southwest direction
as the centimeter emission (1.3, 2, 3.5 and 6 cm), they do not have,
within the error, the same position angle (the millimeter emission
has a position angle of $\sim 61\degr$, while the centimeter emission
has a position angle of $\sim 44\degr$). These results are
consistent with those found very recently by Gibb and Hoare (2007).

The water maser emission toward S140 IRS was in a period of minimum activity
during our VLA observations and we only detect three water maser
features spatially associated with IRS~1A (Figure \ref{IRS1}).
One of these maser features, with redshifted velocity with respect
to the  molecular cloud velocity ($\sim -6.5$~km~s$^{-1}$, Zhou et al.
1993), is located close ($0\farcs1$) to the IRS~1A
continuum emission peak at 1.3~cm, while the other two maser features,
with blueshifted velocity with respect to the molecular cloud velocity,
are located $\sim 0\farcs5$ and $\sim 0\farcs7$ to the southwest of
IRS~1A, respectively (see Fig. \ref{IRS1} and Table \ref{maser}).
Figure 9 from Trinidad et al. (2003) shows that the water
maser emission toward S140 IRS was stronger (up to 46 Jy with at
least seven features) a month before the VLA observations
presented here. The position, velocity, and flux density of the three
detected VLA water maser features are given in Table \ref{maser}.

\section{Discussion}

The elongated morphology of the IRS~1A is evident in the contour maps at 
all centimeter wavelengths (1.3, 2, 3.5, and 6~cm), suggesting a jet-like 
nature (see Figure \ref{IRS1}), similar to other possible radio jets
associated with massive YSOs (e.g., Claussen et al. 1994, Rodr\'{\i}guez
et al. 1994, Torrelles et al. 1997, Hofner et al. 1999, Gibb et al. 2003,
Trinidad et al. 2003, 2005, Curiel et al. 2006; see Hoare 2006 and references 
therein). However, Hoare (2006), based on very detailed 6~cm continuum 
multiepoch observations with MERLIN, has proposed that the continuum
emission is produced by an equatorial wind.  We discuss both of these 
possibilities below.

\subsection{IRS~1A: A Thermal Radio Jet?}
\label{jet}

In order to investigate whether or not IRS~1A is a thermal jet, we have
calculated, following the formalism of Reynolds (1986), the dependence of the
flux density and the deconvolved angular size of the major axis of the jet
with the frequency. For the case of a thermal jet with constant velocity, 
temperature, and
ionization fraction, the dependence of the flux density with the frequency
is $S_\nu \propto \nu^\alpha$ ($\alpha = {1.3-0.7/\epsilon}$; equation 14
of Reynolds 1986), while that for the deconvolved angular size of the major
axis is  $\theta_{maj}\propto \nu^\beta$ ($\beta = {-0.7/\epsilon}$). 
Under this formalism, $\epsilon$ is the power-law index that describes
the dependence of the jet half-width, $w$ (perpendicular to the major
axis of the jet), as a function of the distance from the origin. For a
conical jet, $\epsilon = 1$, then, $\alpha = 0.6$ and $\beta = -0.7$.

From the multiepoch observations at 6~cm (Hoare 2006), it is known
that the flux density of IRS~1A is not variable with time. 
Assuming that the flux density of IRS~1A has not changed significantly
with time also at other wavelengths (the observations were made with a time
span of about 12 years; see Table \ref{observations}),
we roughly estimate a spectral index of $\alpha=0.5\pm0.1$ by using the flux densities at 0.7, 1.3, 2, 3.6, and 6~cm (see Figure \ref{index}).
In addition, we also estimate the dependence of the angular size
of the major axis of IRS~1A with frequency, where the parameter is
$\beta = 0.98\pm0.06$ (see Figure \ref{index}). Both indices, $\alpha$ and $\beta$, seem to be consistent with free-free emission in the wavelength
range 0.7--6 cm, produced either by a conical collimated ionized jet 
($\epsilon$ = 2/3, "standard" collimated jet) or by a conical ionized jet
($\epsilon$ = 1, "standard" spherical jet) (see Table 1 of Reynolds 1986). 
On the other hand, using the flux density of IRS~1A
at 1.3~cm (this work) and that at 6~cm obtained with MERLIN observations
(Hoare 2006), a spectral index of about $0.6\pm0.1$ is obtained, which
is consistent with that estimated using only VLA observations. In this way,
we also note that the flux density and angular diameter of IRS~1A as
measured with the VLA, with an angular resolution of about $0\farcs35$,
and those measured with MERLIN (Hoare 2006), with an angular resolution
of about $0\farcs11$, are consistent within 25--30\%.
This result shows that changes of the angular diameter of the source is
not due to an effect of angular resolution; that is, the larger angular
size measured at low frequencies is not an effect of lower angular resolution
than at higher frequencies.

Under the assumption that IRS~1A is a thermal jet,
we can also estimate the ionized mass-loss ($\dot{M}$) and momentum rate 
 ($\dot{P}$) deposited by IRS~1A into the ambient medium.
Following Reynolds (1986) and using equation 3 given by Beltr\'an
et al. (2001), for a pure hydrogen jet with constant velocity and
ionization fraction, we have
\begin{eqnarray}
\nonumber
   \left( \frac{\dot{M}}{10^{-6}M_\sun yr^{-1}} \right) = 
   0.108\left[ \frac{(2-\alpha)(0.1+\alpha)}{1.3-\alpha} \right]^{3/4}
   \left[ \left( \frac{S_\nu}{mJy} \right) \left( \frac{\nu}{10GHz}
          \right)^{-\alpha} \right]^{3/4}
   \left( \frac{V}{200 km s^{-1}} \right) \\
\times  \left( \frac{\nu_m}{10GHz} \right)^{0.75\alpha-0.45}
   \left( \frac{\theta_o}{rad} \right)^{3/4}
   (\sin i)^{-1/4}
   \left( \frac{d}{kpc} \right)^{3/2}
   \left( \frac{T}{10^4 K} \right)^{-0.075},
\end{eqnarray}
where $\alpha$ is the spectral index, $S_\nu$ is the observed flux density
at frequency $\nu$, $V$ is the terminal velocity of the jet, $\nu_m$ is
the turnover frequency, $\theta_o$ is the opening angle, $i$ is the jet
axis inclination, $d$ is the distance to the source and $T$ is the electron
temperature. The opening angle is approximately estimated using
$  \theta_o = 2 \tan^{-1} (\theta_{min}/\theta_{maj}) $.

Then, assuming a inclination angle of the jet axis relative
the line of sight, $i=60\degr$, a distance of 910~pc, an electron
temperature of $10^4$~K, a lower limit for $\nu_m$ ($=22.2$~GHz), and
a terminal velocity of $\sim 500$~km~s$^{-1}$, we determine
the mass-loss ($\dot{M}$) and momentum rate ($\dot{P}$) deposited by
IRS~1A into the ambient medium as $\sim 9\times10^{-7}$~M$_\sun$~yr$^{-1}$
and $\sim 4.5\times10^{-4}$~M$_\sun$~yr$^{-1}$~km~s$^{-1}$, respectively.
These values are similar to those obtained for the thermal
jets associated with other massive YSOs (e.g Cep A W2 and HH~80). 
On the other hand, using the correlation (momentum rate versus
radio continuum luminosity) given by Anglada (1996) to estimate the momentum
rate, we find a value about two order of magnitude larger than that estimated using the formalism of Reynolds. This result suggests that Anglada's correlation
is valid only for low luminosity objects and cannot be extrapolated to high 
luminosity sources.

\begin{figure} 
\begin{center}
\scalebox{0.35}{
\plotone{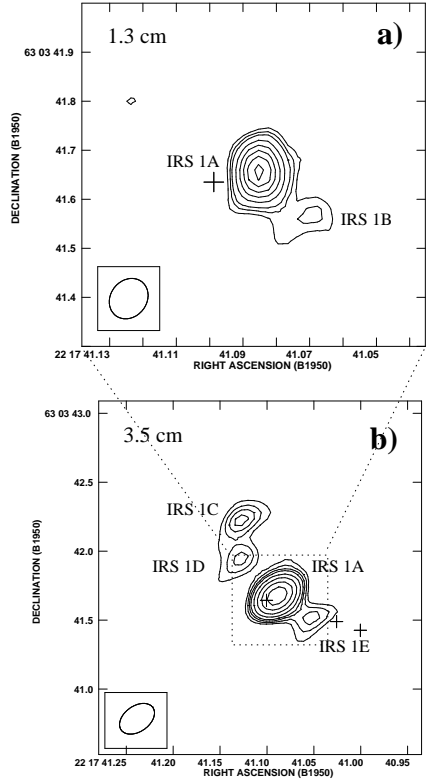}
 }
\end{center}
\caption{Multiple radio continuum sources detected toward S140~IRS~1A
at different wavelengths. {\it a)} 1.3 cm continuum contour map.
Contours are -3,3,4,5,7,9,11,13 and 15 times 230~$\mu$Jy~beam$^{-1}$,
the rms noise of the map. The beam size is $0\farcs09\times0\farcs07$.
{\it b)} 3.5 cm continuum contour map. Contours are -3,3,5,7,9,11,15,20,25
and 30 times 55~$\mu$Jy~beam$^{-1}$, the rms noise of the map. The beam
size is $0\farcs29\times0\farcs20$. The crosses show the position of the water
masers detected in the region. 
} 
\label{multi}
\end{figure}

\begin{figure} 
\plotone{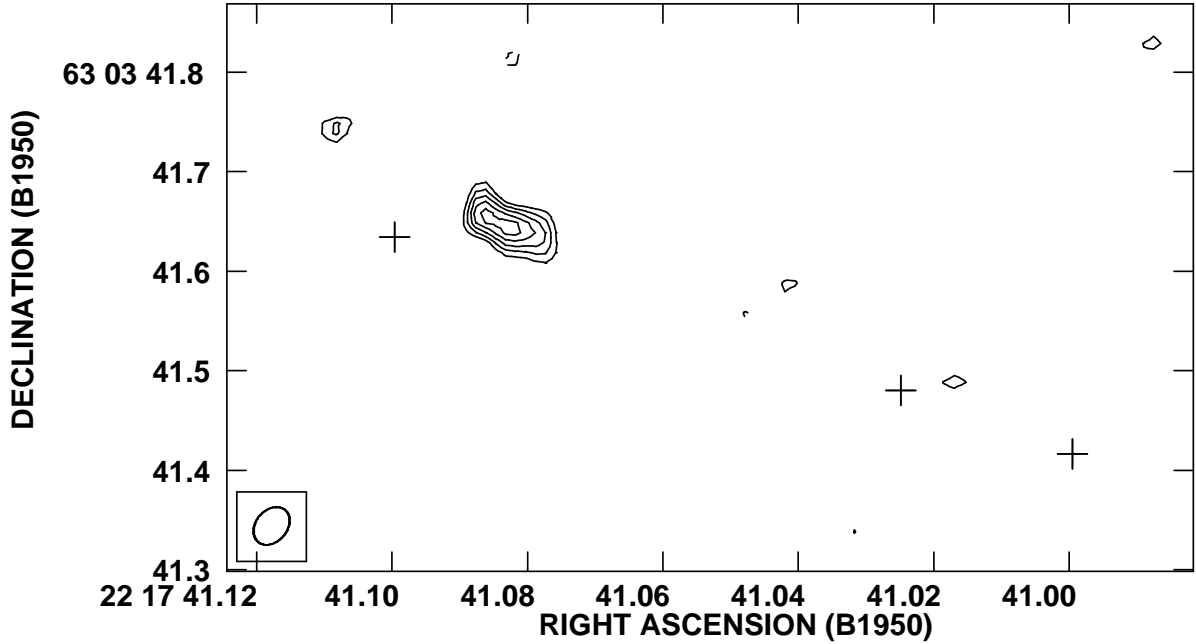}
\caption{Contour map of S140~IRS~1A at 7~mm. Contours are
--3,3,4,5,6 and 7 times 310~$\mu$Jy~beam$^{-1}$,  the rms noise
of the map. The beam size ($0\farcs04\times0\farcs03$)
is shown in the lower left corner. The crosses show the position
of the water masers observed toward the IRS~1A.
} 
\label{Qband}
\end{figure}

\begin{figure} 
\begin{center}
\scalebox{1.0}{
\plotone{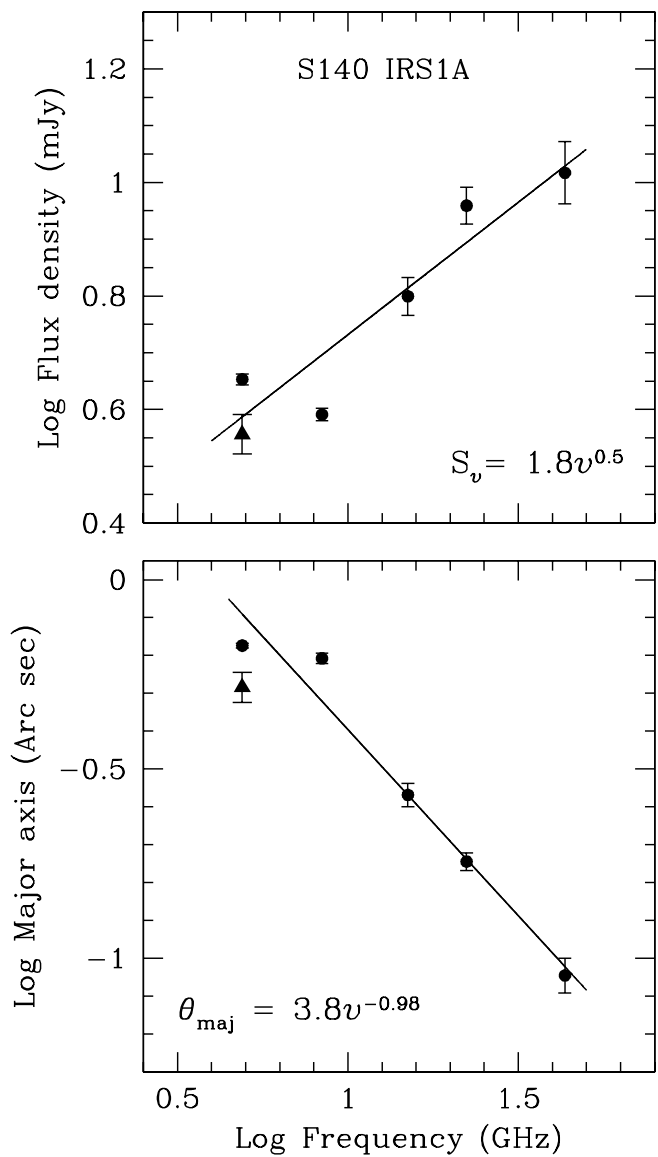}}
\end{center}
\caption{Flux density (top panel) and deconvolved angular size of the major
axis (bottom)
of IRS~1A are plotted as a function of frequency. The solid lines represent
the least-squares fits from which the spectral indices $\alpha$ and $\beta$
are obtained for IRS~1A.
The filled triangles indicate the flux density (upper panel) and the
size of the major axis (lower panel) of IRS~1 as obtained by Hoare (2006).
These values were not used for the least-squares fits.
The flux density, $S_\nu$, is given in mJy, the frequency, $\nu$,
is given in GHz, and the deconvolved size of the major axis, $\theta_{maj}$,
is given in arcsec.
}
\label{index}
\end{figure}

We have also detected four condensations almost aligned along the
major axis of the IRS~1A (see Figure \ref{multi}).
One continuum peak (IRS~1B) is detected at 1.3~cm, while the other
three continuum peaks (IRS~1C, 1D, 1E) are detected at 3.5~cm.
As IRS~1B is detected only at 1.3~cm (although IRS~1B and 1E are located
to the southwest of IRS~1A, both continuum peaks are not spatially
coincident), we are not able to estimate its spectral index, which
could have helped to investigate whether it is a condensation ejected by IRS~1A
or, alternatively, an independent source. The continuum peaks IRS~1D and 1E
are almost symmetrically located with respect to IRS~1A (see Figure 
\ref{multi}) and both condensations have similar peak flux densities,
which could suggest that they are condensations ejected by IRS~1A,
as could be expected for a thermal jet. Multiepoch observations (e.g at 3.5~cm)
to measure possible proper motions of IRS 1D and 1E could test if these are condensations ejected by the jet. On the other hand, the
condensation IRS~1C does not have a counterpart on the symmetrically opposite
side as in the case of the pair formed by 1D and 1E. In addition, IRS~1C appears misaligned with respect to the 1D-IRS~1A-1E system (see Figure \ref{multi}). Then,
if IRS~1C was also ejected by the jet IRS~1A, this could indicate that
IRS~1A is precessing. Alternatively, IRS~1C could be an independent source.
In this way, IRS~1C seems to coincide spatially, within the positional
error, with the clump NE-N detected by Hoare (2006) at 6~cm wavelength.
Using the peak flux densities at 3.5 (this work) and 6~cm (Hoare 2006),
we have roughly estimated a spectral index for
IRS~1C  of $\sim 0.4$, which suggest that the continuum emission
of IRS~1C is being produced by partially-thick thermal free-free emission.
In addition, Hoare (2006) has reported possible proper motion
of the NE-N clump with a tangential velocity of 120~km~s$^{-1}$ moving toward 
the east. However, due to this high value of the velocity, Hoare (2006) opens
the possibility that the proper motion is not due to real physical motion 
but due to a travelling illuminating wave. 

The overall radio continuum properties of IRS~1A indicate that it could
be a thermal jet associated with a high-mass YSO. In addition, the position
angle of the elongated structure at cm wavelengths ($\sim 44\degr$; Table 
\ref{flux}) suggests that IRS~1A could be the driving source of the bipolar 
molecular outflow seen in H$_2$ in the direction $20\degr/200\degr$. Under
this scenario and assuming that the condensation IRS~1C is an
independent source, we speculate that this newly detected source (IRS~1C)
could be the  driving source of the other bipolar molecular outflow
observed in the
direction $160\degr/340\degr$. If this is the case, two close, independent
and almost perpendicular jets could be the driving sources of the two
bipolar outflows observed in the region (see Figure \ref{multi}). 
Perpendicular thermal jets have been observed in low-mass star
forming regions (e.g. HH~111, Reipurth et al. 1999).

In addition, all water maser features detected toward S140~IRS are
associated with IRS~1A, assuming that IRS~1E is not a stellar object
but ejected from IRS 1A (see Figure \ref{IRS1} and \ref{multi}).
The estimated mass for IRS~1A, assuming that the masers are
bound gravitationally, is  $\ge 60$~M$_\sun$, which is greater than the
mass estimated from its luminosity ($\sim 10$~M$_\sun$). This result
indicates that the water maser features are not associated with a
circumstellar disk, instead, the water masers are tracing
unbound motions,
which is consistent with IRS~1A being a thermal jet. 
Further observations with high angular resolution and proper motion
studies will be needed to study in detail the three-dimensional (spatio-kinematical)
distribution of the water masers in the region, which will help to
understand the nature of the IRS~1A and possibly discriminate between
the thermal jet and equatorial wind models (see below).

\subsection{IRS~1A: An Equatorial Wind?}
\label{wind}

Using high angular resolution multiepoch observations at 6~cm toward
IRS~1A, Hoare (2006) proposes that IRS~1A is an equatorial
wind driven by radiation pressure from a central star and inner disk
acting on the gas in the surface layers of the disk, and perpendicular
to the CO bipolar outflow observed in the direction 160$\degr$/340$\degr$.
This scenario is supported by two facts. First, the location and extension perpendicular to the major axis of IRS~1A of a small 
scale monopolar near-IR reflection nebula at the base of the blueshifted
lobe of the molecular outflow, and second, the small outward proper motion
of a continuum clump in the direction 160$\degr$/340$\degr$
as would be expected for a jet.

The results of Hoare are opposed to the picture of the multiple jets described
above for IRS~1A. However, our observational results could be also
consistent with IRS~1A being an equatorial wind. As mentioned above (\S~4.1),
there are two condensations (IRS~1D and 1E) located symmetrically on
opposite sides of the central source, IRS~1A, and
 under the equatorial wind model, IRS~1D and 1E could be tracing
the outer border of the wind within a disk with diameter of $\sim 650$~AU.

Unfortunately, with the present data we cannot discriminate between the
two proposed models for IRS~1A (jet versus equatorial wind). However,
we note that the presence of the outflow in the 20$\degr$/200$\degr$
direction (in addition to the outflow in the 160$\degr$/340$\degr$
direction), requires at least two independent powering YSOs in the
region. In this way, within the equatorial wind scenario IRS~1A would be
the exciting source of the  160$\degr$/340$\degr$ outflow (as proposed
by Hoare 2006), while the new detected source IRS~1C would power the 
20$\degr$/200$\degr$ outflow. 

Additional observations are required to discriminate in a more definitive
way between the equatorial wind and the thermal jet scenarios. Any
detection of large proper motions (hundred of km s$^{-1}$)
in the IRS~1D and 1E condensations in the northeast-southwest direction 
would favor the jet hypothesis.
This test would require sensitive, high angular resolution continuum observations over several years. On the other hand, the study of the high velocity molecular 
gas in the vicinity of the source IRS~1A could reveal the wide angle
outflow expected from an equatorial wind. Any detection, with high
angular resolution observations, of molecular gas correlated with
the radio continuum source could trace the neutral part of the disk
and also provide strong evidence favoring the equatorial wind scenario. 
In addition, as we have mentioned before, proper motion studies of the water
masers observed in this region could be a powerful tool to discriminate between
both models.
On the other hand, with the present theoretical models (e.g. Drew et al.
1998; Lugo et al. 2004), we cannot also discriminate between jet and
equatorial wind models. However, we think 
that in an equatorial wind model the angular diameter might be less
sensitive to the frequency of observation than in an jet model. 
In any case, more detailed theoretical studies will
be necessary to address this important issue.

\section{Conclusions}

We have presented results of 1.3~cm continuum and water maser emission
observations made with the VLA in its A configuration toward the
S140~IRS region. We have also re-analyzed continuum observations at
0.7, 2, 3.5, and 6~cm.

We observed IRS~1A at all wavelengths (0.7, 1.3, 2, 3.5, and 6~cm)
and also detected four new continuum peaks (IRS~1B, 1C, 1D and 1E).
IRS~1B is only detected at 1.3~cm, while IRS~1C, 1D and 1E are
detected at 3.5~cm. IRS~1C, located$\sim 0\farcs6$ to the northeast
of IRS~1A, seems to be an independent source, while IRS~1D and 1E,
located symmetrically with respect to IRS~1A, seem to be condensations
associated with IRS~1A. Under this scenario, IRS~1A is not a single
source, but a possible binary formed by IRS 1A and 1C. These 
two YSOs could explain the excitation of the two large 
bipolar molecular outflows observed toward S140~IRS1.
We have also detected three water masers toward IRS~1 that most probably
are tracing unbound motions and are associated with the bipolar molecular 
outflow in the northeast-southwest direction.

In order to understand the nature of IRS~1A, we have analyzed two scenarios:
the thermal jet model and the equatorial wind model. The elongated morphology
and spectral indices, $\alpha$ and $\beta$, of the continuum emission in the      
0.7 -- 6~cm wavelength range, are consistent with IRS~1A being a thermal jet.
However, a photoevaporated disk (Hoare 2006, Gibb \& Hoare 2007) 
could produce a similar morphology and spectral energy distribution.
Proper motion measurements of the detected
continuum clumps IRS 1D and 1E would help to discriminate between these
two scenarios (jet versus equatorial wind). In addition, proper motion
measurements
of the detected water masers would also be very valuable in this sense.

\acknowledgments
M.A.T. acknowledges the support from CONACyT grant 46157-E.
J.M.T. acknowledges partial financial support from
the Spanish grant AYA2005-08523-C03

\clearpage

\begin{deluxetable}{cccl}
\tablecolumns{4}
\tablewidth{0pc}
\tablecaption{Summary of the continuum observations toward S140~IRS~1}
\tablehead{
\colhead{Wavelength} & \colhead{Observation} & \colhead{HPBW} & \colhead{Reference}  \\ 
\colhead{(cm)} & \colhead{Date}  &  \colhead{(arcsec)} &  \colhead{}\\
}
\startdata
0.7  &  1996 Nov 01  &  $\sim 0.04$  &  VLA data archive (project AH5980)  \\
1.3  &  1999 Jun 29    &  $\sim 0.1$    &  This paper  \\
2    &  1987 Sep 12  &     $\sim 0.1$    &  Schwartz (1989) \\
3.5  &  1992 Nov 24  &  $\sim 0.3$  &  Tofani et al. (1995)  \\
6    &  1987 Sep 12  &     $\sim 0.3$  &  Schwartz (1989) \\
\enddata
 \label{observations}
\end{deluxetable}

\begin{deluxetable}{ccccccc}
\tablecolumns{6}
\tablewidth{0pc}
\tablecaption{Parameters of the continuum sources in S140~IRS~1}
\tablehead{
\colhead{Source} & \colhead{Wavelength} & \multicolumn{2}{c}{Position \tablenotemark{a}} & \colhead{Flux Density} & \colhead{Angular Diameter}  \\ \cline{3-4}
&\colhead{(cm)} & \colhead{$\alpha(1950)$} & \colhead{$\delta(1950)$} & 
\colhead{(mJy)} &  \colhead{} \\
}
\startdata
IRS~1A & 0.7  &  22 17 41.083  &  63 03 41.65  & 10.4$\pm$1.4  &  $0\farcs09 \times 0\farcs02$; P.A. $61\degr$  \\
&1.3  &  22 17 41.083  &  63 03 41.64  &  9.1$\pm$0.7  &  $0\farcs18 \times 0\farcs08$; P.A. $42\degr$    \\
&2    &  22 17 41.085  &  63 03 41.67  &  6.3$\pm$0.5  &  $0\farcs27 \times 0\farcs03$; P.A. $45\degr$  \\
&3.5  &  22 17 41.091  &  63 03 41.70  &  3.9$\pm$0.1\tablenotemark{b} &  $0\farcs62 \times 0\farcs11$; P.A. $46\degr$  \\
&6    &  22 17 41.087  &  63 03 41.69  &  4.5$\pm$0.1  &  $0\farcs68 \times 0\farcs07$; P.A. $41\degr$  \\
IRS~1B & 1.3  &  22 17 41.070  &  63 03 41.57  &  0.92\tablenotemark{c}   &  $<0\farcs08$  \\
IRS~1C & 2    &  22 17 41.136  &  63 03 42.08  &  0.40\tablenotemark{c}  &  $<0\farcs15$  \\
       & 3.5  &  22 17 41.128  &  63 03 42.24  &  0.45\tablenotemark{c}  &  $<0\farcs2$   \\
IRS~1D & 3.5  &  22 17 41.129  &  63 03 41.96  &  0.38\tablenotemark{c}  &  $<0\farcs2$  \\
IRS~1E & 3.5  &  22 17 41.045  &  63 03 41.53  &  0.38\tablenotemark{c}  &  $<0\farcs2$  \\
\enddata
\tablenotetext{a}{Units of right ascension are hours, minutes, and seconds,
and units of declination are degrees, arcminutes, and arcseconds.}
\tablenotetext{b}{This value is lower ($\sim$ 50\%) than that reported by
Tofani et al. (1995).}
\tablenotetext{c}{Peak flux density. The beam size is $0\farcs09 \times 0\farcs07$
at 1.3~cm, $0\farcs16 \times 0\farcs14$ at 2~cm, and $0\farcs28 \times 0\farcs19$
at 3.5~cm.}
 \label{flux}
\end{deluxetable}

\begin{deluxetable}{cccc}
\tablecolumns{4}
\tablewidth{0pc}
\tablecaption{Parameters of the water masers observed toward S140~IRS~1}
\tablehead{
\multicolumn{2}{c}{Position \tablenotemark{a}} & \colhead{$V_{LSR}$}  & 
\colhead{Flux Density}   \\ \cline{1-2} 
\colhead{$\alpha(1950)$} & \colhead{$\delta(1950)$} & \colhead{(km s$^{-1}$)} &
\colhead{(Jy)}  \\ 
}
\startdata
22 17 41.099  &   63 03  41.64   &      7.7   &  0.35 \\
22 17 41.024  &   63 03  41.48   &    -16.0  &  2.29 \\
22 17 40.999  &   63 03  41.42   &    -13.4  &  1.44 \\
\enddata
\tablenotetext{a}{Units of right ascension are hours, minutes, and seconds,
and units of declination are degrees, arcminutes, and arcseconds.}
\label{maser}
\end{deluxetable}

\end{document}